\documentclass[preprints,article,accept,moreauthors,pdftex]{Definitions/mdpi}

\firstpage{1} 
\pubvolume{xx}
\issuenum{1}
\articlenumber{5}
\pubyear{2020}
\copyrightyear{2020}
\history{Received: 5 November 2019; Accepted: 10 January 2020; Published: 15 January 2020}
\updates{yes} 

\Title{On Assessing Driver Awareness of Situational Criticalities: Multi-modal Bio-Sensing and Vision-Based Analysis, Evaluations, and Insights}

\Author{{Siddharth Siddharth} * and Mohan M. Trivedi}

\AuthorNames{Siddharth Siddharth  and Mohan M. Trivedi}

\address[1]{%
Department
of Electrical and Computer Engineering, University of California San Diego, La Jolla, CA  92093, USA; mtrivedi@eng.ucsd.edu\\}

\corres{\hangafter=1 \hangindent=1.05em \hspace{-0.82em}Correspondence: ssiddhar@eng.ucsd.edu}

\abstract{Automobiles for our roadways are increasingly using advanced driver assistance systems. The adoption of such new technologies requires us to develop novel perception systems not only for accurately understanding the situational context of these vehicles, but also to infer the driver's awareness in differentiating between safe and critical situations. This manuscript focuses on the specific problem of inferring driver awareness in the context of attention analysis and hazardous incident activity. Even after the development of wearable and compact multi-modal bio-sensing systems in recent years, their application in driver awareness context has been scarcely explored. The~capability of simultaneously recording different kinds of bio-sensing data in addition to traditionally employed computer vision systems provides exciting opportunities to explore the limitations of these sensor modalities. In this work, we explore the applications of three different bio-sensing modalities namely electroencephalogram (EEG), photoplethysmogram (PPG) and galvanic skin response (GSR) along with a camera-based vision system in driver awareness context. We assess the information from these sensors independently and together using both signal processing- and deep learning-based tools. We show that our methods outperform previously reported studies to classify driver attention and detecting hazardous/non-hazardous situations for short time scales of two seconds. We use EEG and vision data for high resolution temporal classification (two seconds) while additionally also employing PPG and GSR over longer time periods. We evaluate our methods by collecting user data on twelve subjects for two real-world driving datasets among which one is publicly available (KITTI dataset) while the other was collected by us (LISA dataset) with the vehicle being driven in an autonomous mode. This work presents an exhaustive evaluation of multiple sensor modalities on two different datasets for attention monitoring and hazardous events classification.}


\begin{document}

\section{Introduction}
With the development of increasingly intelligent vehicles, it has now become possible to assess the criticality of a situation much before the event actually happens. This makes it imperative to understand the criticality of a situation from the driver's perspective. While computer vision continues to be the preferred sensing modality for achieving the goal of assessing driver awareness, the use of bio-sensing systems in this context has received wide attention in recent times \cite{DrAw_1,DrAw_2,DrAw_3}. Most of these studies used electroencephalogram (EEG) as the preferred bio-sensing modality.

The emergence of wearable multi-modal bio-sensing systems \cite{my_TBME,Biovotion} opened a new possibility to overcome the limitations of individual sensing modalities through the fusion of features from multiple modalities. For example, the information related to driver's drowsiness extracted from EEG (which suffers from low spatial resolution especially when not using a very large number of sensors) may be augmented by the use of galvanic skin response (GSR) which does not suffer from electromagnetic noise (but has a low temporal resolution). 

Driver awareness depends highly on the driver's physiology since different people react differently to fatigue and to their surroundings. This means that one-fit-for-all type of approach using computer vision based on eye blinks/closure etc. might not scale very well across drivers. It is here that the use of bio-sensing modalities (EEG, GSR, etc.) may play a useful role in assessing driver awareness by continuously monitoring the human physiology. The fusion of data from vision-based systems and bio-sensors might be able to generate more robust models for the same. Furthermore, EEG with its higher temporal resolution than other common bio-sensors may prove to be very useful for detecting hazardous vs. non-hazardous situations on short time scales (such as 1--2 s) if such situations do not register in the driver's facial expressions. Additionally, the driver's physiology may provide insights into how they react to various situations during the drive which may have a correlation with the driver's safety. For example, heart-rate variability, which was shown to model human stress~\cite{HRV_Stress} may be used as an indicator of when it is unsafe for a person to drive a vehicle.

Deep Learning has many applications in computer vision-based driver-monitoring \mbox{systems \cite{DL_Driver1,DL_Driver2}.} However, these advances have not translated towards the data from bio-sensing modalities. This is primarily due to the difficulty in collecting very large scale bio-sensing data which is a prerequisite for training deep neural networks. Collecting bio-sensing data on a large scale is costly, laborious, and time-consuming. It requires sensor preparation and instrumentation on the subject before the data collection can be started, whereas for collecting images/videos even a smartphone's camera may suffice without the need to undergo any sensor preparation in most cases.

This study focuses on driver awareness and his/her perception of hazardous/non-hazardous situations from bio-sensing as well as vision-based perspectives. We individually use features from three bio-sensing modalities namely EEG, PPG, and GSR, and vision data to compare the performance of these modalities. We also use the fusion of features to understand if and in what circumstances can it be advantageous. To this end, we present a novel feature extraction and classification pipeline that has the ability to work with real-time capability. The pipeline uses pre-trained deep neural networks even in the absence of very large scale bio-sensing data. To the best of our knowledge, this study is the most comprehensive view of using such widely varying sensing modalities towards assessing driver awareness. Finally, we would like to emphasize that the bio-sensors used in this study are very practical to use in the ``real world'' i.e., they are compact in design, wireless and comfortable to use for prolonged time intervals. This choice of bio-sensors was consciously made so as to bridge the gap between laboratory-controlled experiments and ``real-world'' driving scenarios.

The framework presented in this paper is not just a user study but a complete scalable framework for signal acquisition, feature extraction, and classification that was designed with the intent to work in real world driving scenarios. The framework is modular since we extract the information from each sensor modality separately. Finally, we test two hypotheses in this paper. First, we test if the modalities with low-temporal resolution (but easily wearable) namely PPG and GSR can work as well as EEG and vision modality for assessing driver's attention. Second, we test if (and when) the fusion of features from different sensor modalities boost the classification performance over using each modality independently for attention and hazardous/non-hazardous event classification. In the process of studying these two hypotheses, we extract traditional signal processing-based and deep learning-based features from each sensor modality.

\section{Related Studies}
Driver monitoring for assessing attention, awareness, behavior prediction, etc. was usually done using vision as the preferred modality \cite{DriverAttention_Review,LISA_1,LISA_2}. This is carried out by monitoring the subject's facial expressions and eye-gaze \cite{LISA_4} which are used to train machine learning models. However, almost all such studies using ``real-world'' driving scenarios were conducted during daylight when ample ambient light is present. Even if infra-red cameras are used to conduct such experiments at night, vision modality suffers from occlusion and widely varying changes in illumination \cite{DriverAttention_Review}, both of which are not uncommon in such scenarios. Furthermore, it was shown that EEG can classify hazardous vs. non-hazardous situations over short time periods which is not possible with images/videos \cite{Freiburg_EEG}.

On the other hand, if we focus on the bio-sensing hardware, more than a decade ago, those studies in driving scenarios that used the use of bio-sensing modalities suffered from impracticality in the ``real-world'' situations. This is because the bio-sensors were usually bulky, required wet electrodes, and were very prone to noise in the environment. Hence, the studies carried out with such sensors required wet electrode application and monitors in front of participants with minimal motion \cite{Old_Driving1,Old_Driving2}. In the early years of this decade, such bio-sensing systems gave way to more compact ones capable of transmitting data wirelessly while being more resistant to the noise by better EM (electro-magnetic) shielding and advances in mechanical design. Finally, recent advances have led to the development of multi-modal bio-sensing systems and the ability to design algorithms using the fusion of features from various modalities. This was used for various applications such as in affective computing and virtual reality \cite{My_TAC,EEG_Gaze_VR}.

The use of deep learning for various applications relating to driver safety and autonomous driving systems skyrocketed in the past few years. These studies ranged from understanding driving behavior \cite{DL_Driving_1} to autonomous driving systems on highways \cite{DL_Driving_2} to detecting obstacles for cars \cite{DL_Driving_3} among other applications. All such studies only use vision modality, due to, as mentioned previously, the prevalence of large-scale image datasets. However, the use of ``pre-trained'' neural networks for various applications \cite{VGG_2,my_itsc} may provide a new opportunity. Hence, if bio-sensing data can be represented in the form of an image, it should be possible to use such networks to extract deep learning-based optimal feature representation of the image (henceforth called most significant features) even in the absence of large-scale bio-sensing datasets.

Finally, our system pipeline uses multiple bio-sensing modalities in addition to the vision which is not the case with previous state-of-the-art evaluations done on the KITTI dataset \cite{KITTI}. Our previous work on EEG and visual modality data with the KITTI dataset \cite{my_IV} showed the utility of using these two modalities for driver attention monitoring but does not present a holistic view of multiple bio-sensing modalities with short-time-interval analysis on other datasets. The use of another dataset collected by us while driving a car in autonomous mode with the driver strapped with bio-sensing modalities and holistic comparison of multiple sensing modalities is a chief feature of this research study. \mbox{Also, the previous research} studies generally \cite{Freiburg_EEG,Kitti_Dataset_Study} used a single modality and traditional features (i.e., not based on deep neural networks) for classification. Through our evaluation, we show that we easily beat their results with higher-order features and also evaluate our pipeline on a new driving dataset collected in this study.
 
\section{Research Methods}
In this section, we discuss the various research methods that we employed to pre-process the data and extract features from each of the sensor modalities used in this study. We also show the visualization of sensor data for each sensor modality. This is done by showing the time and/or frequency domain sensor data for the sensor modality as no threshold exists which can distinguish between different mental or psychological states directly for each sensing modality even for a single~subject.

\subsection{EEG-based Feature Extraction}
The cognitive processes pertaining to attention and mental load such as while driving are not associated with only one part of the brain. Hence, our goal was to map the interaction between various regions of the brain to extract relevant features related to attention. The EEG was initially recorded from a 14-channel Emotiv EEG headset at 128 Hz sampling rate \cite{Emotiv}. The EEG channel locations as per the International 10--20 system were AF3, AF4, F3, F4, F7, F8, FC5, FC6, T7, T8, P7, P8, O1, and O2. We used the artifact subspace reconstruction (ASR) pipeline in the EEGLAB \cite{EEGLAB} toolbox to remove artifacts related to eye blinks, muscle movements, line noise, etc. \cite{EEG_ASR}. This pipeline is capable of working in real time and unlike Independent Component Analysis (ICA) \cite{ICA} has the added advantage of being able to remove noise without much loss of EEG data when a very large number of EEG sensors are not present. For each subject, we verified the output from ASR manually as well as observed the algorithm's output parameters to make sure that the noise removal is being performed correctly. Then, we band-pass filtered the EEG data between 4--45 Hz. The band-pass filter was designed so as to capture the theta, alpha, beta, and low-gamma EEG bands i.e., frequency information between 4--45~Hz. On this processed EEG data, we employed two distinct and novel methods to extract EEG features that capture the interplay between various brain regions to map human cognition.

\subsubsection{Mutual Information-Based Features}
\label{sec.3.1.1}
To construct the feature space that can map the interaction of EEG information between various regions of the brain, we calculated the mutual information between signals from different parts of the brain. EEG-based mutual information features were used since they measure the changes in EEG across the various regions of the brain as opposed to power spectrum-based features that are local. This is because unlike a steady-state visual-evoked potential (SSVEP) or a P300 type of EEG response which mostly affects a single brain region, driving is a high-level cognition task and hence multiple systems of the brain are involved: vision, auditory, motor, etc. The mutual information $I(X;Y)$ of discrete random variables $X$ and $Y$ and their instantaneous observation $x$ and $y$ is defined as 

\begin{equation}
I(X;Y) = \sum_{x\in X}\sum_{y\in Y}p(x,y) log\bigg(\frac{p(x,y)}{p(x)p(y)}\bigg)
\end{equation}
where $p(x)$, $p(y)$, and $p(x,y)$ are the probability density function of \emph{X}, \emph{Y}, and joint probability density function of \emph{X} and \emph{Y} respectively \cite{conditional_entropy}. The desired feature of conditional entropy $H(Y|X)$ is related to the mutual information $I(X;Y)$ by

\begin{equation}
I(X;Y) = H(Y) - H(Y|X)
\end{equation}


We calculated the conditional entropy using mutual information between all possible pairs of EEG electrodes for a given trial. Hence, for 14 EEG electrodes, we calculated 91 EEG features based on this measure.

\subsubsection{Deep Learning-Based Features}
\label{sec.3.1.2}
The most commonly used EEG features are the calculation of power-spectrum density (PSD) of different EEG bands. However, these features in themselves do not take into account the EEG-topography i.e., the location of EEG electrodes for a particular EEG band. Hence, we developed a way to exploit EEG-topography for extracting information regarding the interplay between different brain regions. 

Since the 2D spectrum image of the brain PSD has the information about amplitude distribution of each frequency band across the brain, we calculated the PSD of three EEG bands namely theta (4--7 Hz), alpha (7--13 Hz) and Beta (13--30 Hz) for all the EEG channels. The choice of these three specific EEG bands was made since they are the most commonly used bands and are thought to carry a lot of information about human cognition. We averaged the PSD for each band thus calculated over the complete trial. These features from different EEG channels were then used to construct a two-dimensional EEG-PSD heatmap for each of the three EEG bands using bicubic interpolation. These~heat-maps now contained the information related to EEG topography in addition to spectrum density at each of these locations. 

\begin{figure}[H]\centering
\includegraphics[width=15 cm]{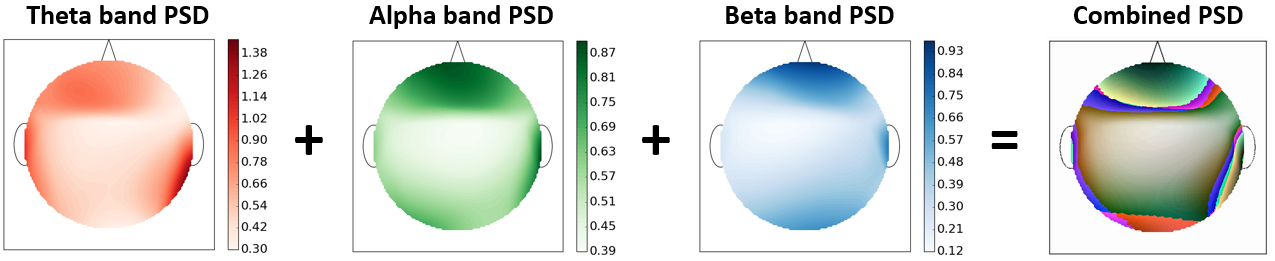} 
\caption{PSD heat-maps of the three EEG bands i.e., theta (\textcolor{red}{{red}}), alpha (\textcolor{green}{green}), and beta (\textcolor{blue}{blue}) EEG bands are added according to respective color-bar range to get combined RGB heat-map image.(Circular outline, nose, ears, and color-bars were added for visualization only. All units are in Watt per Hz.)}
\label{fig:eeg-bands}{} 
\end{figure}

Figure \ref{fig:eeg-bands} shows these 2-D heatmaps for each of the three EEG bands. As can be seen from the figure, we plotted each of the three EEG bands using a single color channel i.e., red, green and blue \cite{RGB_DL}. We then added these three color band images to get a color RGB image containing information from the three EEG bands. The three color band images were added in proportion to the amount of EEG power in the three bands using alpha blending \cite{MatPlotLib} by giving weights to the three individual bands' images by normalizing them using the highest value in the image. Hence, following this procedure we were able to represent the information in the three EEG bands along with their topography using a single color image. The interaction through the mixture of these three colors (thus forming new colors by adding these primary colors) in various quantities was contained the information regarding the distribution of power spectrum density across the various brain regions.

It is notable that we computed PSD in three EEG power bands (theta, alpha, and beta) separately because these three EEG power bands contribute the most towards human cognition. Subsequently, the power in these EEG bands were averaged separately for each band and not together. This process was done so that the RGB colored image with three EEG bands could be constructed using the averaged power of theta, alpha, and beta band separately. It was to execute this complete procedure of generating the RGB colored image with each color band representing the averaged PSD of a particular EEG band that it was not possible to directly average the PSD values between 4--30 Hz.

Since it is not possible to train a deep neural network from scratch without thousands of trials from the EEG data (and no such dataset currently exists in driving scenario), the combined colored image representing EEG-PSD with topography information was then fed to a pre-trained deep learning-based VGG-16 convolution neural network \cite{VGG} to extract features from this image. This network was trained with more than a million images for 1000 object categories using the Imagenet Database \cite{ImageNet}. Specifically, the VGG-16 deep learning network used by us has a total of 16 layers (as shown in Figure 4), the input to the network being 224 $\times$ 224 $\times$ 3 sized colored image and Rectified Linear Unit (ReLU) optimizer as the activation function for all hidden layers. The convolution layers are 3 $\times$ 3 kernel size while the maxpooling layers are 2 $\times$ 2 kernel size and finally there are three layers of the VGG-16 network which are fully connected ones.

Previous research studies \cite{VGG_2,my_itsc} showed that using features from such an ``off-the-shelf" neural network can be used for various classification problems with good accuracy.  Even for the research problems where the neural networks were trained on a different vision-based problem and applied to a totally different application they still worked very well \cite{my_itsc, VGG_3, VGG_4}. This is mostly because the low-level features such as texture, contrast, etc. reflected in the initial layers of the Convolution Neural Network (CNN) are ubiquitous in any type of image. The EEG-PSD colored image was resized to 224 $\times$ 224 $\times$ 3 for input to the network. The last layer of the network classifies the image into one of the 1000 classes but since we were only interested in ``off-the-shelf'' features, we extracted 4096 features from the last but one layer of the network. The EEG features from this method were then combined with those from the previous one for further analysis.

\subsection{PPG-Based Feature Extraction}
PPG measures the changes in blood volume in the microvascular tissue bed. This is done in order to assess the blood flow as being modulated by the heart beat. Using a simple peak detection algorithm on the PPG signal, it is possible to calculate the peaks of the blood flow and measure the subject's heart rate in a much more wearable manner than a traditional electrocardiogram (ECG) system. The PPG signal was recorded using an armband (Biovotion) that measures PPG at a sampling rate of 51.2 Hz.

\subsubsection{HRV and Statistical Time-Domain Features}
\begin{figure}[H] \centering
{\includegraphics[width=\linewidth]{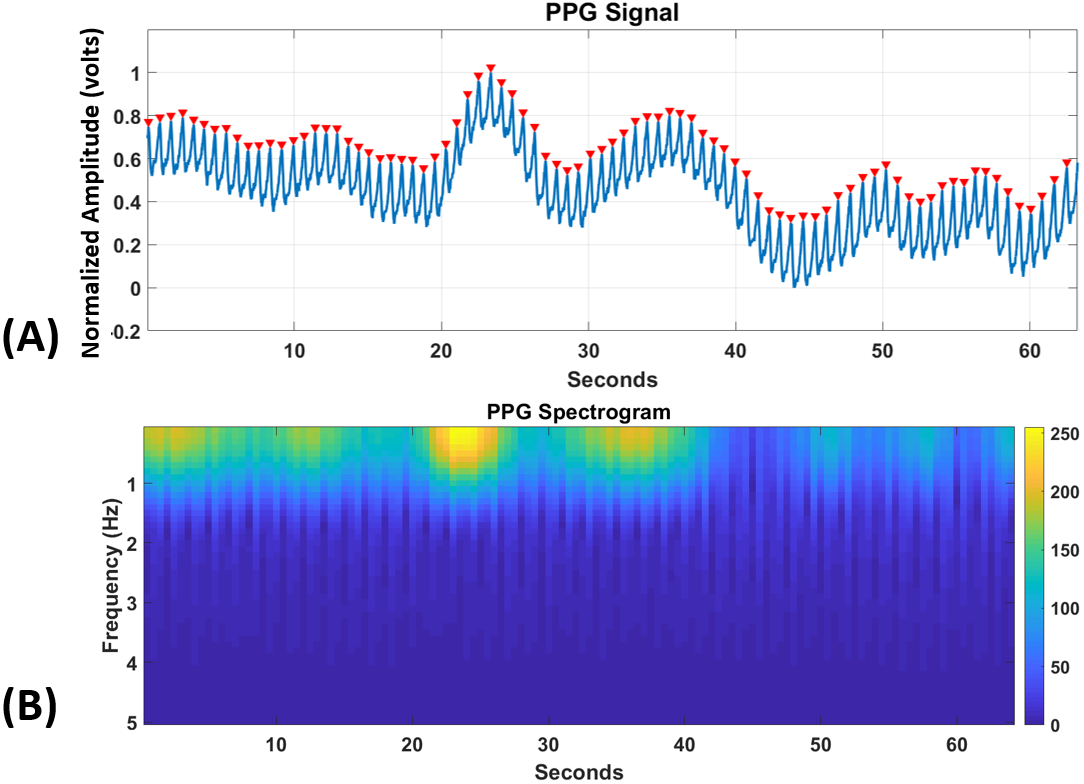}}
\caption{For a trial, PPG signal with peaks (in \textcolor{red}{red}) being detected for the calculation of RRs and HRV (above), and PPG spectrogram (below).}
\label{fig:ppg-plot}
\end{figure}

Heart-rate variability (HRV) has shown to be a good measure for classifying cognitive states such as emotional valence and stress \cite{hrv_only}. HRV is much more robust than heart rate (HR) which changes slowly and generally only correspond to physical exertion. A moving-average filter with a window length of 0.25 s for filtering the noise in the PPG data was first used for each trial. The filtered PPG data so obtained was then scaled between 0 and 1 and subsequently a peak-detection algorithm \cite{peak_detect} was applied to find the inter-beat intervals (RR) for the calculation of HRV. The minimum distance between successive peaks was taken to be 0.5 s to remove any false positives as shown in Figure \ref{fig:ppg-plot}. 

HR is defined as the total number of peaks per minute in the PPG. pNN50 algorithm \cite{pNN50_algorithm} was then used to calculate HRV from RR intervals. To explore the statistics related to the PPG wave itself in time-domain, we calculated six statistical features on the PPG wave as defined in \cite{gsr_features}. These features mapped various trends in the signal by the calculation of mean, standard deviation, etc. at first and subsequent difference signals formed using the original signal.

\subsubsection{Spectrogram Deep Learning-Based Features}
Recent research studies showed promising results by analyzing PPG in the frequency domain for applications such as blood pressure estimation and gender identification  \cite{PPG_Frequency1,PPG_Frequency2}. Despite the low amount of information that is present in the PPG spectrogram as can be seen from Figure \ref{fig:ppg-plot}, we had to use this method since only then we could convert the one-dimensional PPG signal into a two-dimensional image for the computation of higher-order deep learning-based features. 

The frequency range of PPG signals is low and hence we focus only on 0--5 Hz range. Figure~\ref{fig:ppg-plot} shows the generated frequency spectrogram \cite{spectrogram} for this frequency range for the PPG signal in a trial. The different color values generated using the ``Parula'' color-map shows the intensity of the spectrogram at a specific frequency bin. Then, we resized the spectrogram images to feed them to the VGG-16 network (as we did above for the color EEG-PSD images), and after which the 4096 extracted features were extracted from the VGG-CNN network. Time-domain statistical and HRV features from the method above were concatenated with these features for further analysis. 

\subsection{GSR-Based Feature Extraction}
The feature extraction pipeline on the GSR signal was similar to that on the PPG. The same two methods that were applied on the PPG were used for GSR too. Same as PPG, the signals are sampled at 51.2 Hz by the device.

\subsubsection{Statistical Features}
The GSR data were first low-pass filtered with a moving average window of 0.25 s to remove any bursts in the data. Eight features based on the profile of the signal were then calculated. The first two of these features were the number of peaks and the mean of absolute heights of the peaks in the signal. Such peaks and their time differences may prove to be a good measure of arousal. The remaining six features were calculated as in \cite{gsr_features} like the PPG signal above. For this time-series analysis of GSR signal, the features comptuted based on the peaks of GSR  takes into account the Skin Conductance Response (SCR) i.e., the “peaks” of the activity while other features based on the GSR signal profile (by calculating mean and standard deviation of successive differences of the signal) accounts for the Skin Conductance Level (SCL).

\subsubsection{Spectrogram Deep Learning-Based Features}
Since GSR signals change very slowly we focused only on the 0--2 Hz frequency range. We~generated the spectrogram image for GSR in the above frequency range for each trial. This~choice of using low-frequency features was done because similar to PPG, GSR signals change very slowly. We~then sent the spectrogram image to the VGG-16 deep neural network and extract the most significant 4096 features from the same. These features were then concatenated with the features from the time-domain analysis.

\subsection{Facial Expression-Based Feature Extraction}
As discussed above, the analysis of facial expressions has been the preferred modality for driver attention analysis. Hence, our goal is to use this method to compare it against the bio-sensing modalities. Furthermore, most of the research work in this area was done by tracking fixed localized points on the face based on face action units (AUs). Hence, below we show a novel deep learning-based method to extract relevant features from the faces for driver attention and hazardous conditions~detection.

First, we extracted the face region from the frontal body image of the person captured by the camera for each frame. This was done by fixing a threshold on the image size to reduce its extreme ends and placing a threshold of minimum face size to be 50 $\times$ 50 pixels. This resizing was done to remove any false positives and decrease the computational space for face detection. We then used the Viola-Jones object detector with Haar-like features \cite{Viola_Jones} to detect the most likely face candidate.

\subsubsection{Facial-Points Localization-Based Features}
\begin{figure}[H] \centering
{\includegraphics[width=5in]{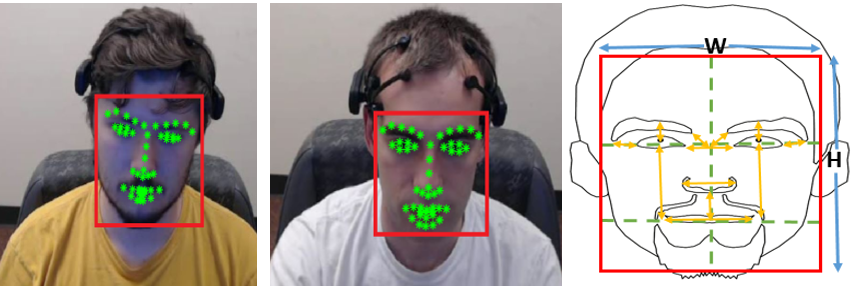}}
\caption{Detected face (marked in \textcolor{red}{red}) and face localized points (marked in \textcolor{green}{green}) for two participants (left and center) in the study, and some of the features (marked in \textcolor{yellow}{yellow}) computed using the coordinates of the face localized points. These features were then normalized using the size of the face in the camera i.e., number of pixels in height (H) and width (W).}
\label{fig:face-features}
\end{figure}

Face action units were used for a variety of applications ranging from affective computing to face recognition \cite{face_AUs}. Facial Action Coding System (FACS) is the most commonly used method to code facial expressions and map them to different emotional states \cite{FACS}. Our goal was to use face localized points similar to the ones used in FACS without identifying facial expressions such as anger, happiness, etc. since they are not highly relevant in the driving domain and short time intervals. The~use of FACS initially involves the identification of multiple facial landmarks that are then tracked to map the changes in facial expressions. We applied the state-of-the-art Chehra algorithm \cite{chehra} to the extracted face candidate region from above. This algorithm outputs the coordinates of 49 localized points (landmarks) representing various features of the face as shown in Figure \ref{fig:face-features}. The choice of this algorithm was done because of its ability to detect these landmarks through its pre-trained models and hence not needing training for any new set of images. These face localized points were then used to calculate 30 different features based on the distances such as between the center of the eyebrow to the midpoint of the eye, between the midpoint of nose and corners of the lower lip, between the midpoints of two eyebrows, etc. and the angles between such line segments. To remove variations by factors such as distance from the camera and face tilt, we normalized these features using the dimensions of the face region. All these features were calculated for individual frames, many of which make a trial. Hence, to map the variation in these features across a trial (which may directly correspond to driver's attention and driving condition) we calculated the mean, 95th percentile (more robust than maximum), and the standard deviation of these 30 features across the frames in the trial. In this manner, we computed 90 features based on face-localized points from a particular trial.

\subsubsection{Deep Learning-Based Features}
For the extraction of deep learning-based features, we used the VGG-Faces deep learning network instead of VGG-16 \cite{VGG_Faces}. This was done to extract features more relevant to faces since the VGG-Faces network was trained on more than 2.6 million face images from more than 2600 people rather than on various object categories in the VGG-16 network. We sent each face region part to the network and extracted the most significant 4096 features. To represent the changes in these features across the trial i.e., across the frames, we calculated the mean, 95th percentile, and standard deviation of the features across the frames in a trial. We then separately analyzed the features from this method to those from the traditionally used face-localized points-based method from above to compare the two.

\subsection{Assessing Trends of EEG/Face Features Using Deep Learning}
\label{sec.3.5.}
\begin{figure}[H] \centering
{\includegraphics[width=\linewidth, height = 1.7in]{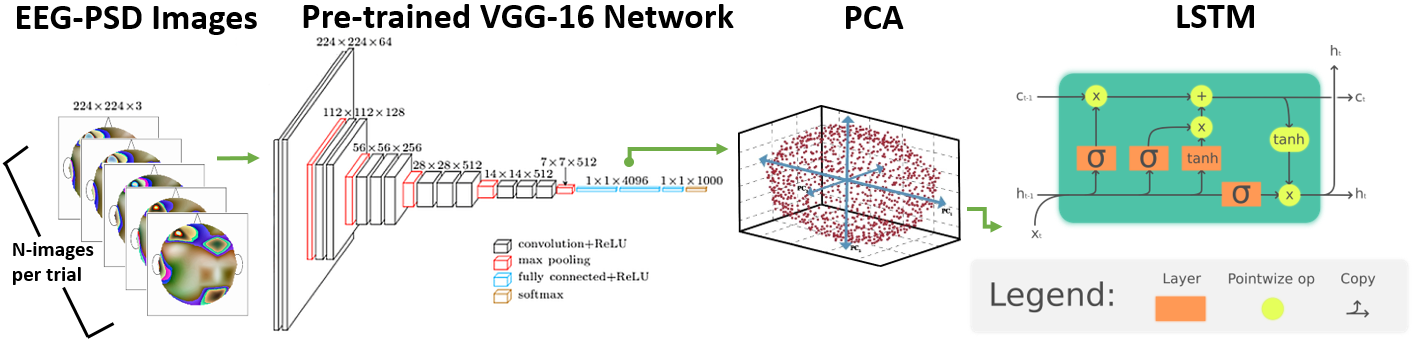}}
\caption{Network architecture for EEG-PSD trend-based Deep Learning method.}
\label{fig:eeg_method_6}
\end{figure}

The EEG features discussed in Section \ref{sec.3.1.2} above were computed over the whole trial such as by generating a single EEG-PSD image for a particular trial. This is a special case when the data from the whole trial is being averaged. Here, we propose a novel method to compute the trend of EEG features i.e., their variation in a trial based on deep learning. To compute features with more resolution we generated multiple EEG-PSD images for successive time durations in a trial. We generated one image per second of the data for driver attention analysis and used 30 images for every second for a 2-s incident classification analysis as detailed below in the Quantitative Analysis section. Figure~\ref{fig:eeg_method_6} shows the network architecture for this method. The EEG-PSD images were generated for multiple successive time durations in a trial each of which was then sent to the VGG-16 network to obtain 4096 most significant features. Similarly, this process was done for conditional entropy features by calculating this over multiple periods in a trial rather than once on the whole trial. We then used principal component analysis (PCA) \cite{PCA} to reduce the feature size to 60 to save computational time in the next step. These 60 $\times$ \emph{N} (\emph{N} = number of successive time intervals) features were then sent as input to a Long Short Term Memory (LSTM) network \cite{LSTM}.

\textls[-5]{In particular, the use of LSTM was motivated by extracting information from sensor modalities with higher temporal resolution. Example: For the EEG and modality, the extraction of deep-learning features without LSTM was being done by representing the whole trial with a single 2D EEG image. This~was a sort of average power-spectrum density image for the whole trial and thus had a bad temporal resolution. Instead, we later calculated one such power-spectrum density image for every second to observe for the 2D image patterns change from second to second. This high temporal resolution was modeled using LSTM which further increases the accuracy by using these shifting patterns.}

The LSTM treats each of these features as a time-series and was trained so as to capture the trend in each of them for further analysis. This method could only be applied when the time duration of the trials is fixed since the length of each time series should be the same. Hence, we applied this method only in the trials used for detecting hazardous/non-hazardous situations and on EEG and face i.e., vision sensor modalities. However, since the analysis of such situations was done on a short time intervals basis, we could not use this method for PPG and GSR modalities since they take a few seconds to react physiologically to the situation.

\section{Dataset Description}
\begin{figure}[H]
\centering
\includegraphics[width=5in, height=2in]{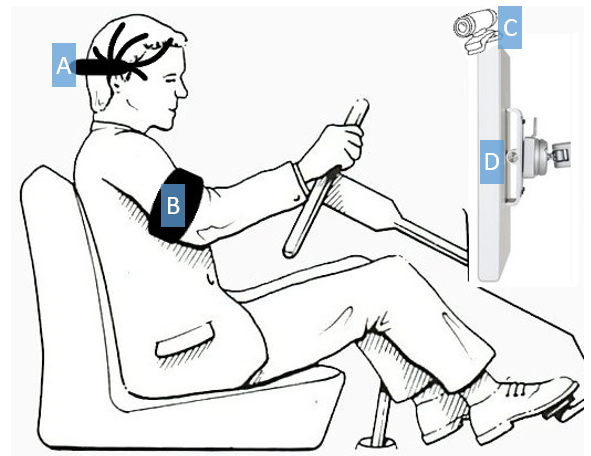} 
\caption{Experiment setup for multi-modal data collection. (A) EEG Headset, (B) PPG and GSR armband, (C) External camera, and (D) Driving videos displayed on the screen. The subject sits with her/his arms and feet on a driving simulator with which s/he interacts while watching the driving~ videos.}
\label{fig:exp_setup}
\end{figure}

Figure \ref{fig:exp_setup} shows the experimental setup for data collection with driving videos used as the stimulus in our dataset. Twelve participants (most of them in their 20 s with two older than 30 years) based in San Diego participated in our study. The participants were comfortably seated equipped with EEG headset (Emotiv EPOC) containing 14 EEG channels (sampling rate of 128 Hz.) and an armband (Biovotion) for collecting PPG and GSR (sampling rate of 51.2 Hz). This EEG headset was chosen since it is easily wearable and does not require the application of electrode gel. This made the headset conform more closely to real-world applications such as in the driving context. However, these~advantages came at the cost of two limitations, namely, lower sampling rate and fewer EEG channels as compared to bulky EEG headsets used in the laboratory. The positioning of the GSR sensor was however sub-optimal since we do not place the sensor at the palm or the feet. This choice was driven by the practicality of data collection in the driving scenario since users interact with multiple vehicle modules from their palms and feet during driving. The facial expressions of the subject were recorded using a camera in front of him/her. The participants were asked to use a driving simulator which they were instructed to control as per the situation in the driving stimulus. For example, if there was a ``red light'' or ``stop sign'' at any point in a driving stimulus video, the participants should press and hold the brake.
\unskip
\begin{figure}[H] \centering
{\includegraphics[width=\textwidth]{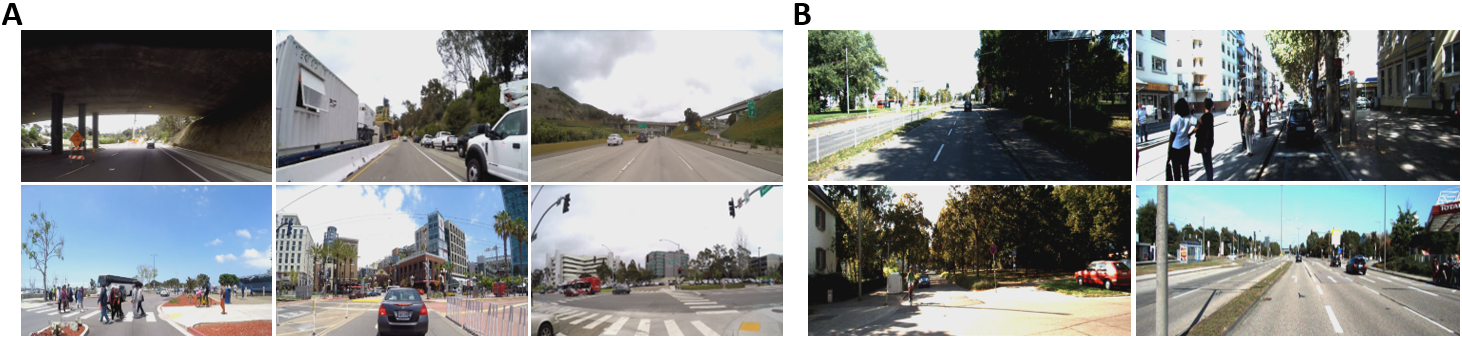}}
\caption{{Various image instances} with varying illumination conditions and type of road (street, single-lane, highway, etc.) from (\textbf{A}) LISA Dataset and (\textbf{B}) KITTI Dataset.} {} 

\label{fig:dataset_image}
\end{figure}

For consistency between our work and other previous studies \cite{Freiburg_EEG,Kitti_Dataset_Study}, we used 15 video sequences from the KITTI dataset \cite{KITTI}. These previous research studies  used the same dataset but without detailing the exact image sequences used to generate the video sequences. Thus, the video sequences in this study chosen based on external annotation by two subjects to judge them based on potential hazardous events in them. These video sequences ranged from 14 to 105 s. These video sequences in the dataset were recorded at 1242 $\times$ 375 resolution at 10 frames-per-second (fps). We resized the videos to 1920 $\times$ 580 to fit the display screen in a more naturalistic manner. However, video sequences from the KITTI dataset suffer from three limitations namely low resolution, low fps, and a few sequences of driving on highways. Additionally, since the images in the KITTI dataset were captured at 10 fps it may produce steady-state visual-evoked potential (SSVEP) effect in EEG \cite{SSVEP}. This is undesirable since we only focus on driver attention and hazardous/non-hazardous events analysis and SSVEP might act as a noise in the process.

Hence, we also collected our own dataset of 20 video sequences containing real-world driving data on freeways and downtown San Diego, California. This dataset was collected using our LISA-T vehicle testbed in which a Tesla Model 3 is equipped with 6 external facing GoPro cameras. It is also to be noted that while capturing these videos the vehicle was in the autonomous driving mode making LISA dataset the first of its kind. The cameras were operating at 122 degrees field-of-view which is very representative of the human vision. Furthermore, we presented these video sequences on a large screen (45.9 inches diagonally) at a distance of a meter from the participants to model a real-world driving scenario. These video sequences ranged from 30 to 50 s in length and were shown to the participants with 1920 $\times$ 1200 resolution at 30 fps. External annotation was done to classify parts of the video sequences from both datasets into hazardous/non-hazardous events. For example, an~event where a pedestrian suddenly appears to cross the road illegally was termed hazardous whereas an event where a stop sign can be seen from a distance and the vehicle's speed is decreasing was termed non-hazardous. External annotation was performed to classify every video sequence into how attentive the driver ought to be in that particular sequence. In Table \ref{table-dataset-params}, we catalog the different datasets and features that we used in the evaluation of our proposed pipeline. In Fig. \ref{fig:dataset_image}, we show some examples of driving conditions from KITTI and LISA dataset.

\begin{table}[H]
\centering
\caption{Dataset-related parameters}
\label{table-dataset-params}
\begin{tabular}{ccc}
\toprule
\centering
\textbf{Dataset} & \textbf{KITTI} & \textbf{LISA}\\
\midrule
\textbf{Number of video sequences} & 15 & 20 \\
\textbf{Time Duration per video (s)} & 14--105 & 30--50 \\
\textbf{Frames per second} & 10 & 30 \\
\textbf{Video resolution} & 1920 $\times$ 580 & 1920 $\times$ 1200  \\\midrule
\textbf{Sensor Modality} & \textbf{Traditional Features} & \textbf{Deep-learning Features}\\
\midrule
\textbf{EEG} & 96 (Conditional Entropy) & 4096 (EEG-PSD 2-D spectrum image) \\
\textbf{PPG} & 7 (HRV and Statistical) & 4096 (PPG spectrogram image) \\
\textbf{GSR} & 8 (Statistical) & 4096 (GSR spectrogram image) \\
\textbf{Face video} & 30 (Face AUs-based) & 4096 (Face image-based)\\
\bottomrule
\end{tabular}
\begin{tabular}{ccc}
 \multicolumn{1}{c}{\footnotesize Table showing the various parameters pertaining to datasets and features used in evaluation.}
\end{tabular}
\end{table}

\section{Quantitative Analysis of Multi-Modal Bio-Sensing and Vision Sensor Modalities}
\label{sec.5.}
In this section, we present the various singular modality and multi-modal evaluation results for driver attention analysis and hazardous/non-hazardous instances classification. First, the videos in both datasets were externally annotated by two annotators for low/high driver attention required. For~example, the video instances where the car is not moving at all were characterized as low attention instances whereas driving through narrow streets with pedestrians on the road were labeled as instances with high driver attention required. Hence, among the 35 videos (15 from KITTI dataset and 20 from LISA dataset), 20 were characterized as requiring low-attention and 15 as high-attention ones.

\begin{figure}[H] \centering
{\includegraphics[width=\linewidth]{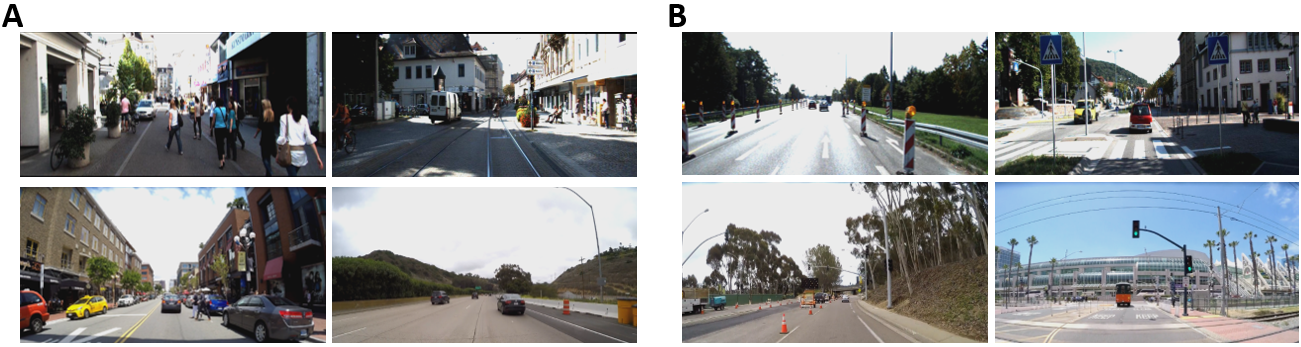}}
\caption{(\textbf{A}) Examples of 2-seconds incidents classified as hazardous. Examples include pedestrians crossing the street without a crosswalk while the ego vehicle is being driven and another vehicle overtaking suddenly. (\textbf{B}) Examples of 2-seconds incidents classified as non-hazardous. Examples include stop signs and railway crossing signs. For each category, the top images are from KITTI dataset whereas the bottom images are from LISA dataset.}
\label{fig:incidents-examples}
\end{figure}

Second, 70 instances, each two-second long were found in the videos and were characterized as hazardous/non-hazardous. Figure \ref{fig:incidents-examples} presents some examples of instances from both categories. As~an example, a pedestrian suddenly crossing the road ``unlawfully'' or a vehicle overtaking suddenly represents hazardous events whereas ``red'' traffic sign at a distance and a pedestrian at a crossing with ego vehicle not in motion are examples of non-hazardous events. Among the 70 instances, 30~instances were labeled as hazardous whereas rest were labeled as non-hazardous. Hence, the goal was to classify such instances in a short time period of two seconds using the above modalities. Since PPG and GSR have low temporal resolution and do not reflect changes in such short time intervals, we used only facial features and EEG for hazardous/non-hazardous event classification.

For each modality, we first used PCA \cite{PCA} to reduce the number of features from the above algorithms to 30. We then used extreme learning machines (ELM) \cite{ELM} for classification. The choice of using ELM over other feature classification methods was driven by previous studies that showed how ELM performs better for features derived from bio-signals \cite{ELM_1, ELM_2}. These features were normalized between $-$1 and 1 across the subjects before training. A single hidden layer ELM was used with a triangular basis function for activation. For the method with trend-based temporal EEG and face feature data, we used two-layer LSTM with 200 and 100 neurons in respective layers instead of ELM for classification. The LSTM network's training was done using stochastic gradient descent with a momentum (SGDM) optimizer. We performed leave-one-subject-out cross-validation for each case. This meant that the data from 11 subjects (385 trials) were used for training at a time and the classification was done on the 35 trials from the remaining 12th subject. This choice of cross-validation was driven by two factors. First, this method of cross-validation is much more robust and less prone to bias than models such as leave-one-sample-out cross-validation that constitutes training data from all the subjects at any given time. Second, since the data contained 420 trials only as opposed to thousands of trials for any decent image-based deep-learning dataset, it does not make sense to randomly divide such a small number of trials to training, validation and test sets since it might introduce bias by uneven division across trials from individual subjects. 

Both of the feature classification methods i.e., LSTM-based and ELM-based were used independently for feature classification with labels. When a higher temporal resolution was taken into consideration i.e., trends in a series of EEG-PSD images, then the LSTM-based method was used for feature classification. This is because now the features vary as a time series for each trial and ELM cannot be used for such a time-series-based classification. The ELM-based method was performed for the other case i.e., the case when high temporal resolution data (multiple data point features for each trial) were not present. The data from the complete trial was represented by a single (non-varying in time) value for each feature.

As pointed out by a reviewer of our manuscript, accuracy in itself may not be the best metric for evaluation in 2-class classification problems \cite{50percent_classification}. This is because 50\% is not an optimal threshold in itself since the number of training and testing samples in the two-class classification problem in most datasets (including ours) are not exactly equal and this threshold only works for an infinite number of data samples. We used accuracy as a metric in consonance with previous research studies using the same in the field of multi-modal bio-sensing research \cite{DEAP, Mahnob-HCI} while we also note the area under the curve (AUC) as another metric in our analysis. AUC is a more robust classification metric for 2-class classification problems and was used by previous research studies in driver attention analysis~\cite{Freiburg_EEG, Kitti_Dataset_Study}.

\subsection{Evaluating Attention Analysis Performance}
In this section, we evaluate single and multi-modality performance for assessing the driver's attention across the video trials. For all the four modalities, the features as defined above were calculated for data from each video trial. The ELM-based classifier was then trained based on each video trial divided into one of the two classes representing low-attention and high-attention required by the driver.

\subsubsection{Singular Modality Analysis}
\begin{figure}[H] \centering
{\includegraphics[width=5in]{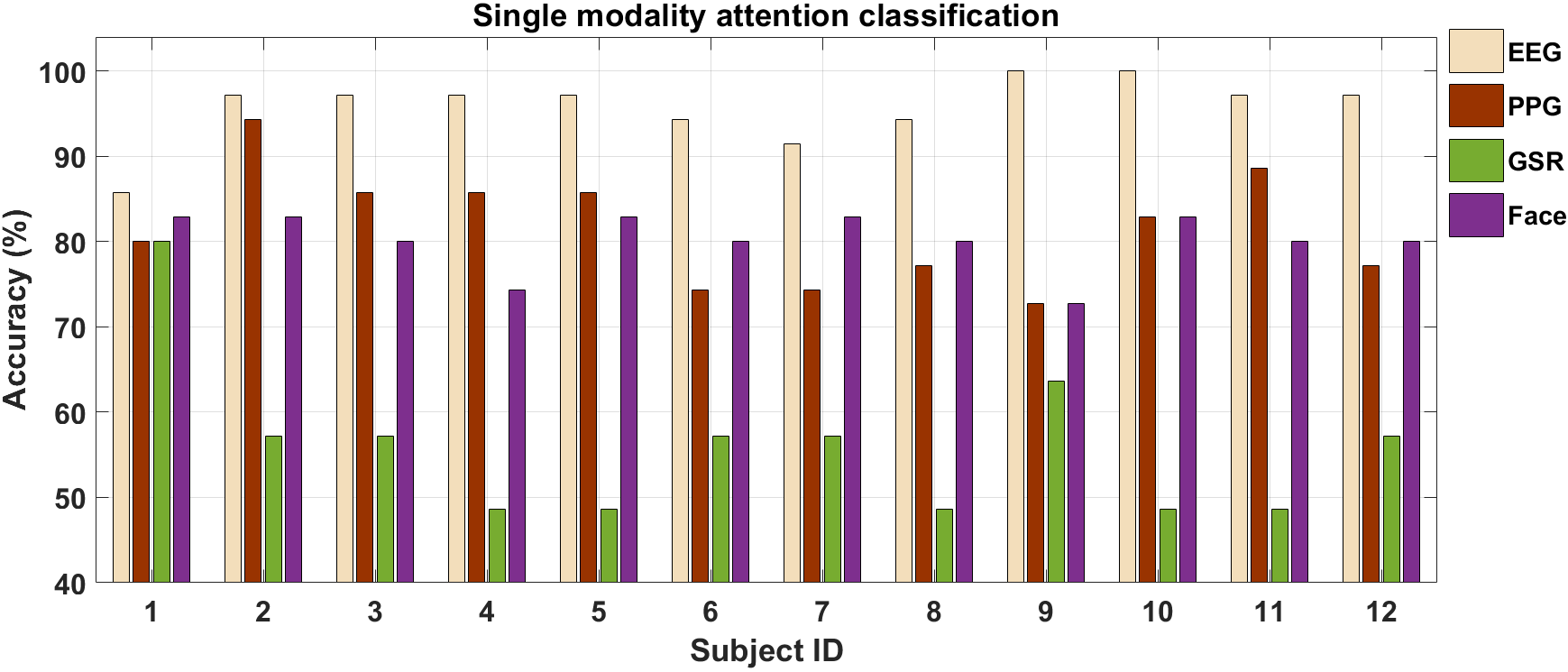}}
\caption{Single modality classification performance for driver attention analysis.}
\label{fig:attention-single-modality}
\end{figure}

To compare the performance among the different modalities, the number of neurons in the hidden layer was set to 170 for each of the modality. In Figure \ref{fig:attention-single-modality} we show these results. Clearly, EEG performs the best among the four modalities for driver attention classification. The average classification accuracy for EEG, PPG, GSR, and face-videos were $95.71 \pm 3.95\%, 81.54 \pm 6.67\%, 56.02 \pm 3.04\%$, and $80.11 \pm 3.39\%$, respectively. The AUC (area under the curve) for the above four cases were $0.84 \pm 0.01, 0.83 \pm 0.02, 0.71 \pm 0.19$, and $0.79 \pm 0.18$ respectively. We also performed statistical t-test on our observations. The p-values for the above four classification cases were $10^{-5}$, $10^{-6}$, $10^{-3}$, and $10^{-6}$ respectively. Hence, GSR performs only at about chance level whereas on average PPG and face videos perform equally well. Thus, we see that the sensor modalities with good temporal resolution i.e., EEG and vision perform better or at least as good as the ones with low temporal resolution (PPG and GSR) thus evaluating our first hypotheses. We performed pairwise t-test for all six pair combinations of the above four signal modalities and the p-values were less than 0.05 for all cases. Thus, the statistical tests showed the results to be statistically significant. We also see that for all the subjects except one, EEG's~classification accuracy is above 90\% while for three modalities (EEG, GSR, and vision) the standard deviation in performance across the subjects is not too high.

\subsubsection{Multi-Modality Analysis}
\begin{figure}[H] \centering
{\includegraphics[width=5in]{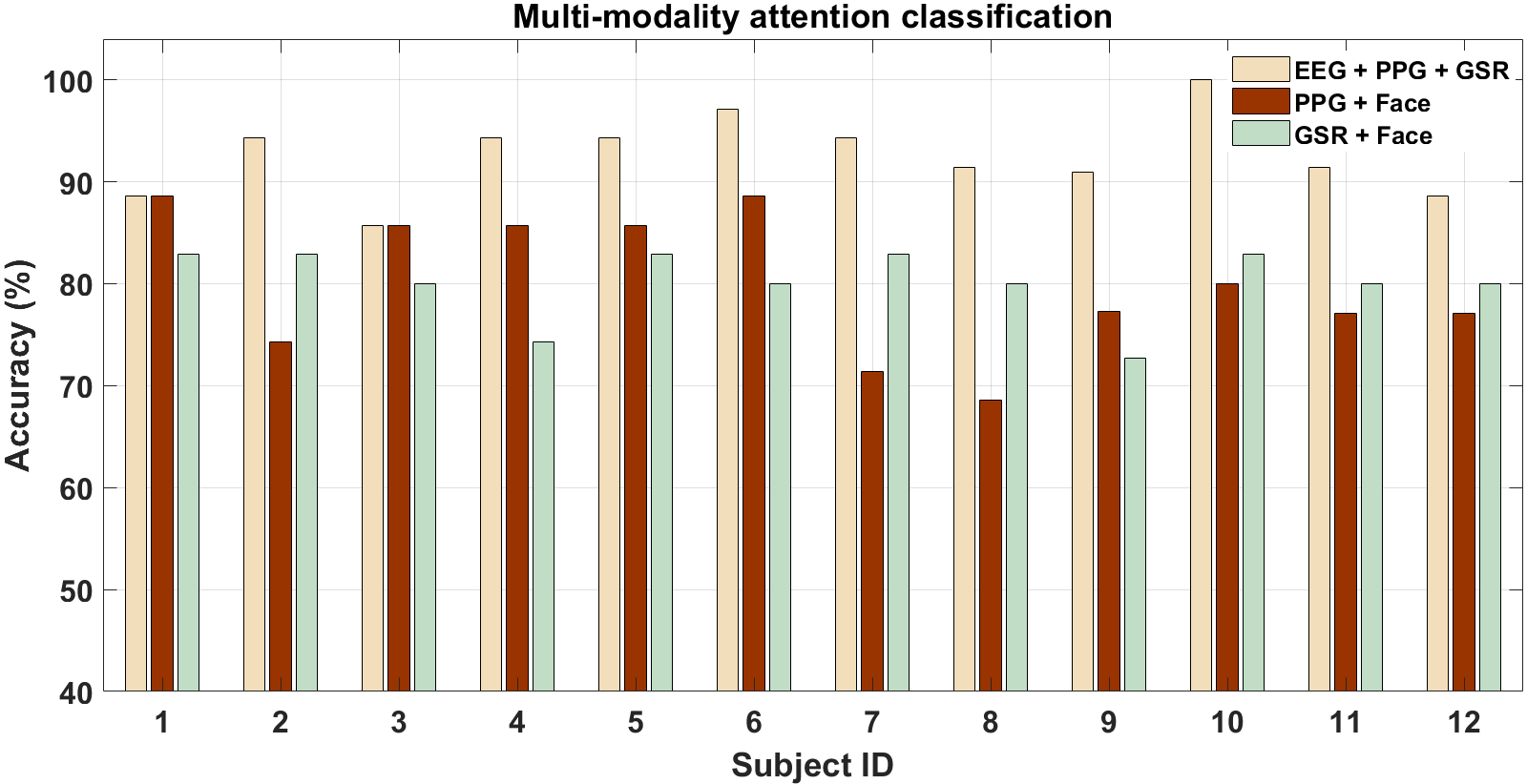}}
\caption{Multi-modality classification performance for driver attention analysis.}
\label{fig:attention-multi-modality}
\end{figure}

Since EEG performs best among the four modalities by far, we do not expect much further increase in classification accuracy while combining it with other modalities that perform much worse. Figure \ref{fig:attention-multi-modality} shows that on combining EEG with PPG and GSR there is no increase in the performance across the subjects (it might be that for a few subjects this is not the case). On the contrary, when~the features from the low-performing (and poor temporal resolution) modalities i.e., PPG and GSR are combined with EEG, the performance is not as good as EEG alone for most of them. The mean accuracy across all the subjects were $92.58 \pm 3.96\%, 80.11 \pm 3.39\%$, and $80.01 \pm 6.78\%$ for the three cases respectively, all of which were significantly above the chance accuracy. The AUC for the above three cases were $0.85 \pm 0.01, 0.80 \pm 0.03$, and $0.80 \pm 0.01$ respectively. The p-values for the above four classification cases were $10^{-6}$, $10^{-6}$, and $10^{-3}$ respectively. To compare the different signal combinations, we also performed pairwise t-test for the above cases. The p-values of pairwise t-test for multi-modal attention classification were $10^{-6}$ between (EEG + PPG + GSR) and (GSR +Face) cases and 0.9 between (GSR + Face) and (PPG + Face). Finally, we also performed pairwise t-test analysis between multi-modality and single-modality cases and found that the p-values between all four singular modalities (EEG, PPG, GSR, and Face) and the three multi-modality cases mentioned above were less than 0.05. These p-values thus denote that not all signal combination cases between multi-modality cases are statistically significant in a pairwise manner while those between singular and multi-modality cases were statistically significant. Hence, we see that it is not always beneficial to use features from multiple sensor modalities. For most of the subjects and modalities, the fusion of features does not perform better at all and hence may not be advantageous in this case. We think that this is because of the vast difference in the performance of each modality when used independently, based on the subject's physiology. This leads to an increase in performance for some of the subjects but not for all. However, for all combinations of sensor modalities we see that the accuracy values were as good or better than using individual sensor modalities which proves our hypotheses about the performance improvement that could be gained by the use of multiple sensor modalities.

\subsection{Evaluating Hazardous/Non-hazardous Incidents Classification}
In this section, we present the results of the evaluation of the modalities over very short time intervals (2 s) pertaining to hazardous/non-hazardous driving incidents as shown in Figure \ref{fig:incidents-examples}. Since~GSR and PPG do not provide such a fine temporal resolution, we do not use these modalities for this evaluation. This is because GSR changes very slowly i.e., take more than a few seconds to vary and PPG for a very short time period such as 2 s would mean only 2--4 heartbeats which are not enough for computing heart-rate or heart-rate variability. Previous studies to assess human emotions using GSR and PPG on the order of multiple seconds (significantly greater than two seconds hazardous incident evaluation for driving context) \cite{Emotion_Study}. Also, it is not possible for the subjects to tag the incidents while they are participating in the driving simulator experiment and hence these incidents were marked by the external annotators as mentioned above in Section \ref{sec.5.}.

\subsubsection{Single-Modality Analysis}
\label{sec.5.2.1}
\begin{figure}[H] \centering
{\includegraphics[width=5in]{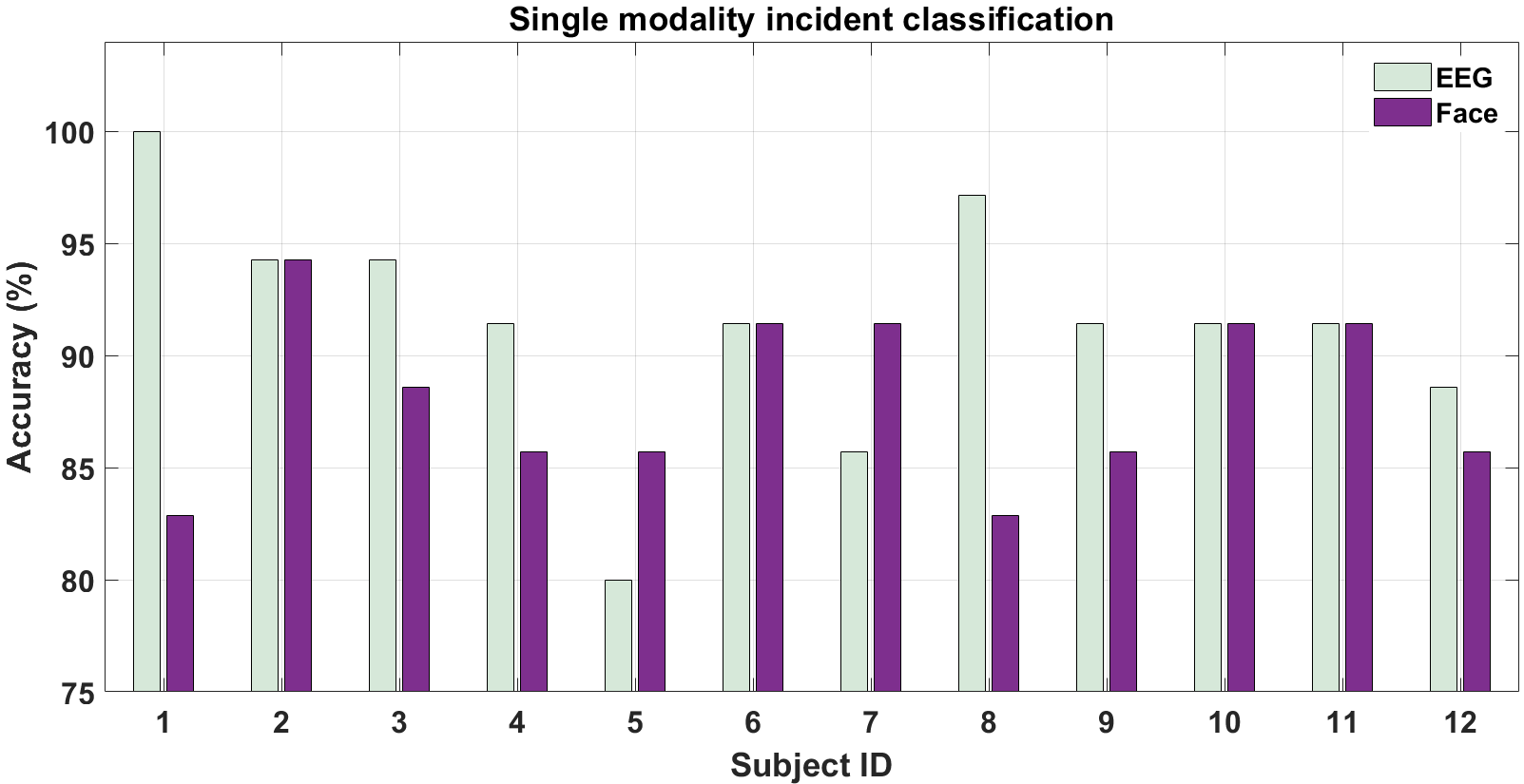}}
\caption{Single-modality classification performance for driver attention during hazardous/non-hazardous incidents.}
\label{fig:incidents-single-modality}
\end{figure}

Figure \ref{fig:incidents-single-modality} shows the results for classifying hazardous/non-hazardous incidents using EEG and face-expression features. As we can see from the figure, the accuracy for both modalities for all the subjects is well above chance level (50\%). The inter-subject variability for different sensor modalities can also been visualized from the above figure. For example, EEG outperforms face-based features for half of the participants but not for the other half. This bolsters our earlier argument: it may be so that bio-sensing modality (here EEG) may outperform vision modality depending on the user's physiology. This variation in results is natural since some people tend to be more expressive with their facial expressions while on the other hand, the ``perceived hazardousness'' of a situation varies across subjects. The mean accuracy among subjects were $91.43 \pm 5.17\%$ and $88.10 \pm 3.82\%$ for EEG-~and face-based features respectively. The AUC for these two cases were $0.85 \pm 0.02$ and $0.84 \pm 0.02$ respectively. The p-values for the above two classification cases were $10^{-5}$ and $10^{-5}$ respectively. Finally, we also performed statistical pairwise t-test on the above two sensor modalities and the p-values was found to be $0.06$. Thus, statistical significance was found within pairwise EEG and Face sensor modalities. Since~the evaluation was done on 2-s time intervals i.e., without a lot of data we note that such a high mean accuracy for both modalities was only possible due to using deep learning-based features in addition to the traditional features for both modalities. This is further substantiated by the fact that we used an EEG system with a much lesser number of channels than such previous studies using EEG \cite{Freiburg_EEG}.  We show that using such deep learning features our method outperforms the previous results for EEG on the KITTI dataset \cite{Freiburg_EEG, Kitti_Dataset_Study} in a similar experimental setup with hazardous/non-hazardous event classification. Specifically, our single-modality approach for both EEG (AUC 0.85) and Face-videos (AUC 0.84) outperform on both datasets the previous best result (AUC 0.79) shown only on the KITTI dataset using EEG alone in \cite{Freiburg_EEG}.

\subsubsection{Multi-Modality Analysis}
\begin{figure}[H] \centering
{\includegraphics[width=5in]{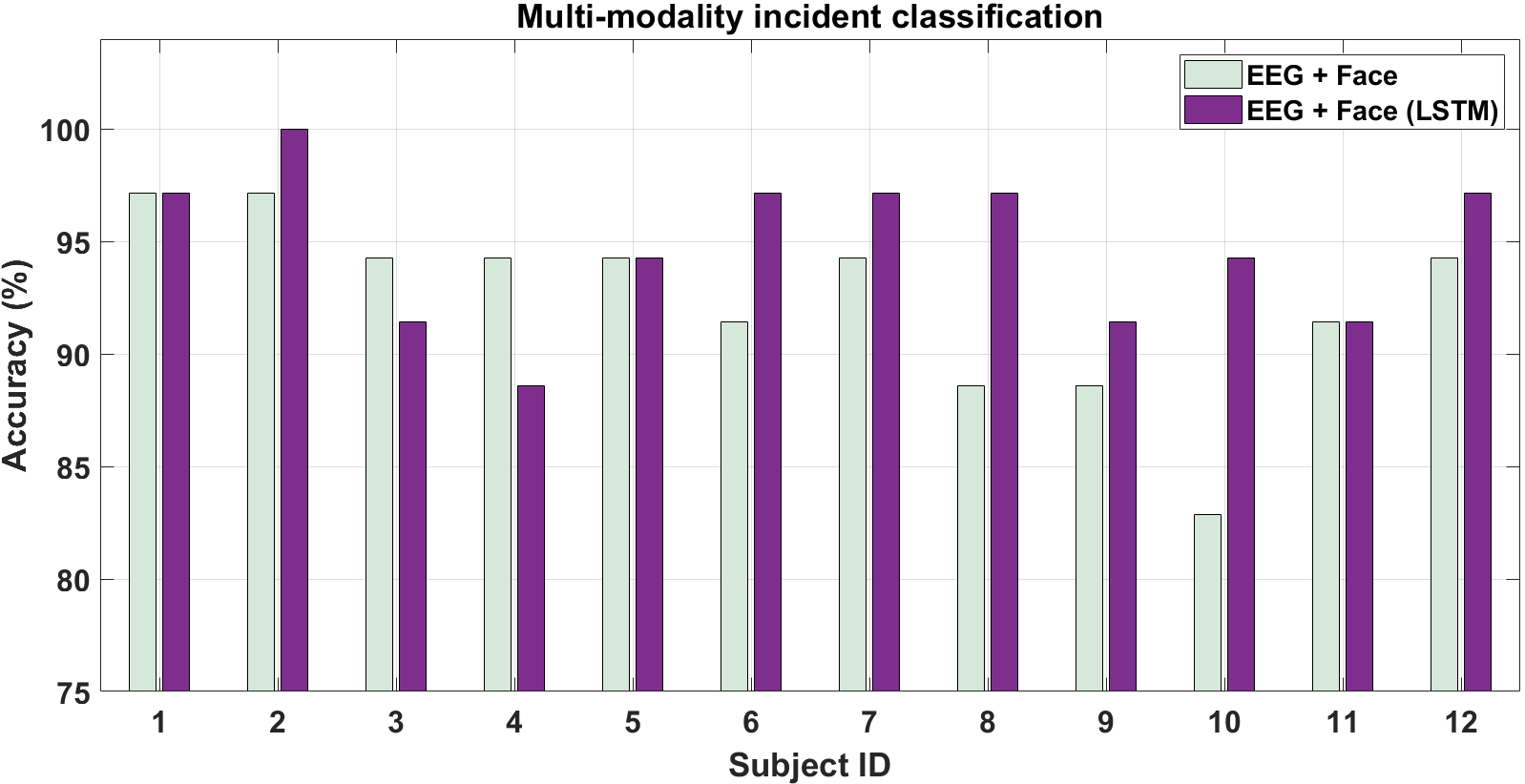}}
\caption{Multi-modality classification performance for driver attention during hazardous/non-hazardous incidents.}
\label{fig:incidents-multi-modality}
\end{figure}

In this section, we present the results for classifying hazardous/non-hazardous incidents combining features from EEG and face modalities. We do this in two ways. First, directly combining the features from single modality analysis by concatenating them. Second, as mentioned in Section \ref{sec.3.5.}, we use an LSTM classifier over the features from both modalities calculated for every frame in the 2-s long sequence. The trend of these features is then fed to LSTM for training. Figure \ref{fig:incidents-multi-modality} shows the results of the two approaches. As is clear from the figure, combining the features from the two modalities may or may not increase the performance further compared to using individual modalities shown earlier in Figure \ref{fig:incidents-single-modality}. However, we note that on taking the trend of features i.e., increased temporal resolution into account, the performance of combining the modalities increases further for most of the subjects. This~can be seen from Figure \ref{fig:incidents-single-modality} where using LSTM-based method, the accuracy increases across the subjects. The average accuracy across subjects being $92.38 \pm 4.10\%$ and $94.76 \pm 3.41\%$ respectively are also more than for singular modality analysis. The AUC for these two cases were $0.84 \pm 0.01$ and $0.88 \pm 0.01$ respectively. The p-values for the above two classification cases were $10^{-6}$ and $10^{-7}$ respectively. We also performed the pairwise t-test for the above multi-modal cases. The~pairwise p-vaue was 0.10 for driving hazardous/non-hazardous incident classification between (EEG + Face) and (EEG + Face (LSTM)) cases and thus show that unlike individual modality cases, the pairwise statistical analysis was not significant for the above pair of signal modality combinations. Finally, we also performed statistical pairwise t-test evaluation between singular sensor modalities from Figure~\ref{fig:incidents-single-modality} and multiple modalities from Figure \ref{fig:incidents-multi-modality}. The p-values between the two singular sensor modality cases (EEG and Face) and two multi-modality cases were less than 0.05. Thus, statistical significance was observed between singular and multi-modality sensor combinations as well. In~conclusion, to use multiple modalities with high temporal resolution (EEG and vision) may prove to be best when computing features over a short time duration with their trend (though it will involve more computational power). 

\begin{table}[H]
\centering
\caption{Single vs. Multi-Modality Performance Evaluation.}
\label{table-performance-evaluation}
\begin{tabular}{lcc}
\toprule
\centering
\textbf{Modality} & \textbf{Attention Analysis} & \textbf{Incident Analysis}
\\\midrule
\textbf{EEG} & $95.71 \pm 3.95\%$ & $91.43 \pm 5.17\%$ \\
\textbf{Faces} & $80.11 \pm 3.39\%$ & $88.10 \pm 3.82\%$ \\
\textbf{EEG + Faces} & $95.10 \pm 3.62\%$ & $92.38 \pm 4.10\%$  \\
\textbf{EEG + Faces (LSTM)} & --- & $94.76 \pm 3.41\%$ \\ \bottomrule
\end{tabular} 
\bigskip
\begin{tabular}{@{}c@{}} 
\multicolumn{1}{p{\textwidth -.88in}}{\footnotesize The average accuracy across subjects using EEG and face-based features for driver attention analysis over the whole video and 2-second hazardous/non-hazardous incident classification. EEG features generally outperform Face features for both cases. Using LSTM i.e., better temporal resolution also increases the accuracy. LSTM could not be used in attention analysis since the duration of the videos varies widely among the datasets.}
\end{tabular}

\end{table}
\vspace{-18pt}

Since EEG and Face modalities can be used in short-time intervals, in Table \ref{table-performance-evaluation} we show the mean accuracy across subjects for using EEG and faces separately and combining them for the two types of analysis done above. We can see that the performance of EEG combined with faces can be better than when either modality is used independently for hazardous incident analysis when using features from the LSTM i.e., trend over the changes in features. However, adding multiple modalities together without using trend-based LSTM analysis may not prove much beneficial. This answers our second hypotheses by showing that it is beneficial to use a fusion of the modalities if both modalities have a good temporal resolution so as to extract short-duration features over them to map the trend. Thus, connecting dots with the single-modality analysis above in Section~\ref{sec.5.2.1}, we can say that multi-modality boosts performance over using individual modalities for hazardous/non-hazardous incident classification (as it did for driver attention analysis) while further improvement in performance is observed by using higher temporal resolution using LSTMs.

\section{Concluding Remarks}
The use of multiple bio-sensing modalities combined with audio-visual ones is rapidly expanding. With the advent of compact bio-sensing systems capable of collecting data during real-world tasks such as driving, it is natural that this research area will gather more interest in the coming years. This~work evaluated multiple bio-sensing modalities with the vision modality for driver attention and hazardous event analysis. We also presented a pipeline to process data from individual modalities by being able to use pre-trained convolution neural networks to extract deep learning-based features from these modalities in addition to traditionally used ones. In this process, we were able to compare the performance of the modalities against each other while also combining them. 

Specifically, as suggested by the academic editor of our manuscript, we also performed Analysis of Variance (ANOVA) test between all singular modality cases from Figure \ref{fig:attention-single-modality} and multi-modality from Figure \ref{fig:attention-multi-modality}, and found the F-statistic to be 0.15. This showed that the between group variance was much lesser than the within group variances. We think that this is due to the fact that within each group there are multiple subjects whose varying physiology accounts for the wide difference in the variances in each sensor modality's evaluation. This is also visible from the two figures in which for most modalities, the accuracy values across the modalities vary a lot between subjects. Similarly, we~performed ANOVA test for the singular modality cases from Figure \ref{fig:incidents-single-modality} and multiple modality cases from Figure \ref{fig:incidents-multi-modality}, and found the F-statistic to be 0.82. This shows that again the between group variance was lesser than the within group variances though not as much as in the driver attention classification evaluation. We think that again this is because human physiology varies widely across subjects and much more data collection needs to be done in the second phase of our study to understand this variation as described below.

As one of the reviewers of our manuscript pointed out, we could not capture the driver-specific baseline for each bio-sensing modality in our evaluation. In fact, we did try to model the driver-specific baseline on our end but could not do so with the present data (~30 min per subject) that we collected in this study. The temporal dynamics of bio-sensors especially EEG are hard to model for such short time-intervals. We also think that having only twelve subjects limited our capability of studying the driver baselines and cluster them into particular categories for analysis. Thus, we are planning to conduct a larger study and perhaps with the real-world driving scenario i.e., driving an actual automobile for the study. The current manuscript acts as a base toward that direction since we were able to develop signal processing and feature extraction pipelines for all sensor modalities. We are sure that with more number of subjects we should be able to better model the driver baselines by categorizing them into clusters. Our ultimate aim in the long-term would be to construct individual models of each driver. This may need a lot of driving data (tens of hours) for every driver with tens of drivers in the study.

As the next step to this preliminary study, we would also collect data in the future in more complex and safety-critical situations from ``real-world" driving. The current system for data collection, noise removal, feature extraction, and classification works in real time but the data tagging for hazardous/non-hazardous events is still manual. This makes the relevance of our current system restricted to gaming, attention monitoring, etc. in driving simulators. Hence, we would like to make a model based on computer vision that can automatically predict hazardous/non-hazardous events during a real-world drive instead of manual tagging so that our system can be deployed in the ``real-world'' drives. We would also like to devise a real-time feedback system based on the driver's attention to verify our pipeline's performance during the real-world driving scenario.

\vspace{6pt} 



\authorcontributions{conceptualization, S.S. and M.M.T.; methodology, S.S. and M.M.T.; software, S.S.; validation, S.S. and M.M.T.; formal analysis, S.S.; investigation, S.S. and M.M.T.; writing---original draft preparation, S.S.; writing---review and editing, S.S. and M.M.T.; supervision, M.M.T.}

\funding{This research received no external funding.}

\acknowledgments{We would like to thank Tzyy-Ping Jung for helping us in revising the manuscript and providing helpful suggestions regarding the same. We would also like to thank our colleagues at the LISA Lab for helping us in the data collection for this study.}

\conflictsofinterest{The authors declare no conflict of interest.} 

\reftitle{References}





\end{document}